\documentstyle[aps,prl,twocolumn, graphics, mytitle,epsf]{revtex}

\begin{document}
\title {Search for the
$\bf K_L \rightarrow \pi^0 \pi^0 e^+ e^-$ Decay in the KTeV Experiment}


\author{
A.~Alavi-Harati$^{12}$,
T.~Alexopoulos$^{12}$,
M.~Arenton$^{11}$,
K.~Arisaka$^2$,
S.~Averitte$^{10}$,
R.F.~Barbosa$^{7,**}$,
A.R.~Barker$^5$,
M.~Barrio$^4$,
L.~Bellantoni$^7$,
A.~Bellavance$^9$,
J.~Belz$^{10}$,
D.R.~Bergman$^{10}$,
E.~Blucher$^4$,
G.J.~Bock$^7$,
C.~Bown$^4$,
S.~Bright$^4$,
E.~Cheu$^1$,
S.~Childress$^7$,
R.~Coleman$^7$,
M.D.~Corcoran$^9$,
G.~Corti$^{11}$,
B.~Cox$^{11}$,
A.R.~Erwin$^{12}$,
R.~Ford$^7$,
A.~Glazov$^4$,
A.~Golossanov$^{11}$,
G.~Graham$^{4}$,
J.~Graham$^4$,
E.~Halkiadakis$^{10}$,
J.~Hamm$^1$,
K.~Hanagaki$^{8}$,
S.~Hidaka$^8$,
Y.B.~Hsiung$^7$,
V.~Jejer$^{11}$,
D.A.~Jensen$^7$,
R.~Kessler$^4$,
H.G.E.~Kobrak$^{3}$,
J.~LaDue$^5$,
A.~Lath$^{10}$,
A.~Ledovskoy$^{\dagger}$$^{11}$,
P.L.~McBride$^7$,
P.~Mikelsons$^5$,
E.~Monnier$^{4,*}$,
T.~Nakaya$^{7}$,
K.S.~Nelson$^{11}$,
H.~Nguyen$^7$,
V.~O'Dell$^7$,
R.~Pordes$^7$,
V.~Prasad$^4$,
X.R.~Qi$^7$,
B.~Quinn$^{4}$,
E.J.~Ramberg$^7$,
R.E.~Ray$^7$,
A.~Roodman$^{4}$,
S.~Schnetzer$^{10}$,
K.~Senyo$^{8}$,
P.~Shanahan$^7$,
P.S.~Shawhan$^{4}$,
J.~Shields$^{11}$,
W.~Slater$^2$,
N.~Solomey$^4$,
S.V.~Somalwar$^{10}$,
R.L.~Stone$^{10}$,
E.C.~Swallow$^{4,6}$,
S.A.~Taegar$^1$,
R.J.~Tesarek$^{10}$,
G.B.~Thomson$^{10}$,
P.A.~Toale$^5$,
A.~Tripathi$^2$,
R.~Tschirhart$^7$,
S.E.~Turner$^2$,
Y.W.~Wah$^4$,
J.~Wang$^1$,
H.B.~White$^7$,
J.~Whitmore$^7$,
B.~Winstein$^4$,
R.~Winston$^4$,
T.~Yamanaka$^8$,
E.D.~Zimmerman$^{4}$\\
(KTeV Collaboration)}

\address{\vspace{0.1in}
\noindent $^{1}$University of Arizona, Tucson, Arizona 85721 \\
$^{2}$University of California at Los Angeles, Los Angeles, California 90095\\
$^{3}$University of California at San Diego, La Jolla, California 92093\\
$^{4}$The Enrico Fermi Institute, The University of Chicago, Chicago, Illinois
60637\\
$^{5}$University of Colorado, Boulder Colorado 80309\\
$^{6}$Elmhurst College, Elmhurst, Illinois 60126\\
$^{7}$Fermi National Accelerator Laboratory, Batavia, Illinois 60510\\
$^{8}$Osaka University, Toyonaka, Osaka 560 Japan\\
$^{9}$Rice University, Houston, Texas 77005\\
$^{10}$Rutgers University, Piscataway, New Jersey 08855\\
$^{11}$The Dept. of Physics and Institute of Nuclear and Particle
Physics, University of Virginia, Charlottesville, Virginia 22901\\
$^{12}$University of Wisconsin, Madison, Wisconsin 53706\\ }

\date{\today}

\myabstract{
The recent discovery of a large CP violating asymmetry in 
$K_L\rightarrow\pi^+\pi^-e^+e^-$ mode has prompted us to seach for 
the associated $K_L\rightarrow\pi^0\pi^0e^+e^-$ decay mode
in the KTeV-E799 experiment at Fermilab.  In $2.7\times 10^{11}$ $K_L$ decays, 
one candidate event has been observed with an expected background of 0.3 
event, resulting in an upper limit for 
the $K_L\rightarrow\pi^0\pi^0e^+e^-$ branching ratio 
of $6.6\times 10^{-9}$ at the 90\% confidence level.  \\ \\ 
\vspace{0.25in}
\noindent
PACS numbers: 13.25.Es, 13.30.Ce, 14.40.Aq}


\maketitle 

The discovery in the KTeV experiment of the
$K_L\rightarrow\pi^+\pi^-e^+e^-$ decay mode~\cite{BR} later confirmed
by KEK-E162 and CERN-NA48~\cite{CBR} together with the observation of
a large CP violating asymmetry~\cite{CP} in the angular distribution
between the $e^+e^-$ and $\pi^+\pi^-$ planes, has generated interest
in searching for $K_L\rightarrow\pi^0\pi^0e^+e^-$.  This letter
details the first attempt to detect this mode.

The $K_L\rightarrow\pi^+\pi^-e^+e^-$ decay can proceed at tree level
via several amplitudes~\cite{CP2}.  The main processes are a) the CP
violating bremsstrahlung process in which the $K_L$ decays into
$\pi^+\pi^-$ where one of the pions radiates a photon which internally
converts into an $e^+e^-$ pair, and b) the CP conserving direct decay
of the $K_L$ into a $\pi^+\pi^-$ pair plus an M1 photon. Other smaller
amplitudes include c) the CP violating direct decay of the $K_L$ into
a $\pi^+\pi^-$ plus an E1 photon and d) the ``charge radius''
amplitude describing the $K_L\rightarrow K_S$ transition by emission
of a virtual photon followed by the CP conserving decay
$K_S\rightarrow\pi^+\pi^-$.  This process is similar to the
regeneration of $K_L\rightarrow K_S$ by scattering from atomic
electrons, both of which are proportional to the mean squared charge
radius of the neutral kaon.  In contrast, the
$K_L\rightarrow\pi^0\pi^0e^+e^-$ decay, because the pions are neutral,
has no bremsstrahlung amplitude.  In addition, the direct emission of
a M1 photon or an E1 photon is suppressed since gauge invariance and
Bose statistics require a $\pi^0\pi^0$ angular momentum of $l\geq
2$~\cite{Lee}.

A consequence of the suppression of the direct M1 and E1 emission
processes and the absence of a bremsstrahlung process is that the
$K_L\rightarrow\pi^0\pi^0e^+e^-$ branching ratio is expected to be
considerably smaller~\cite{CP3} than the
$K_L\rightarrow\pi^+\pi^-e^+e^-$ branching ratio measured to be
$3.2\times 10^{-7}$~\cite{BR}.  The $K_L\rightarrow\pi^0\pi^0e^+e^-$
is predicted to be dominated by the CP conserving ``charge radius''
amplitude with additional contributions from the CP conserving E2 and
CP violating M2 direct emission amplitudes.  Theoretical estimates of
the $K_L\rightarrow\pi^0\pi^0e^+e^-$ branching ratio vary considerably
from a vector meson dominance model estimate of $2.0\times
10^{-8}$~\cite{CP4} and chiral perturbation theory calculations of
$1\times 10^{-10}$~\cite{CP5} and $2.3\times 10^{-10}$~\cite{CP3}.

An observation of $K_L\rightarrow\pi^0\pi^0e^+e^-$ would provide an
indirect measurement of the $K_L\rightarrow\pi^0\pi^0\gamma$ branching
ratio, a mode which is difficult to detect directly because of the
serious backgrounds~\cite{pipig,Barr}.  Observation of this mode would
also allow investigation of chiral perturbation theory contributions
to O($p^4$) and O($p^6$).  The extra contributions of the full chiral
perturbation calculations beyond that of the charge radius amplitude
of Ref.~\cite{CP3} can complicate the extraction of this
amplitude~\cite{CP3,CP5,CP6}.

A search for the $K_L\rightarrow\pi^0\pi^0e^+e^-$ decay mode has been
conducted by the KTeV collaboration using the 1997 E799 data set.  In
the E799 spectrometer configuration, two almost parallel neutral beams
passed through the fiducial decay region of the KTeV spectrometer 95
m$\leq z \leq$ 158 m downstream of a BeO production target.  The beams
were produced by $(3-5)\times 10^{12}$ 800 GeV/c protons per minute
delivered in 20 second spills incident at an angle of 4.8 mrad on a
BeO target. The composition of the beams was mainly $K_L$'s and
neutrons with a small admixture of $\Lambda$'s, and $\Xi$'s.  An
integrated total of $2.7\times 10^{11}$ $K_L$ decays took place in the
fiducial region while the trigger was live. The $K_L$ energies varied
between 20 and 200 GeV with a mean of 70 GeV.

The KTeV spectrometer components which were used for the analysis of
these decays consisted of four stations of drift chambers, two
upstream and two downstream of an analysis magnet which provided a
$p_t$ kick of 0.205 GeV/c for momentum determination of charged
tracks.  Downstream of the last drift chamber, in succession, were a
charged particle trigger hodoscope which provided the first level
trigger requirement of two charged tracks and a 3100 element pure CsI
calorimeter for photon and electron id and energy
determination. Surrounding the decay region and the elements of the
detector were veto counters that detected charged and neutral
particles that were not in the acceptance.  The complete KTeV E799
spectrometer configuration is described in Ref.~\cite{BR}

The data were collected with a trigger which required two charged
tracks in the drift chambers, four or more showers with energies
$\geq$ 1 GeV and total energy in the CsI calorimeter be greater than
30 GeV and no significant energy be present in any segment of the veto
counters surrounding the decay volume and spectrometer elements.

Events were selected off line that had six energy clusters in the CsI,
two of which were associated with the $e^+e^-$ pair.  The $K_L$ decay
vertices were determined by minimizing the $\chi^2$ of a fit which
included the differences between the invariant mass of the secondary
particles from the $\pi^0$ decays ($\gamma\gamma$ or $e^+e^-\gamma$ in
the case of modes with $\pi^0$'s undergoing Dalitz decay as discussed
below) and $\pi^0$ mass together with the distance between the vertex
and the $e^{\pm}$ tracks. The $\chi^2$ incorporated the measured
uncertainties of the energies and positions of the photon clusters in
the CsI and the uncertainties of the $e^{\pm}$ track position due to
the drift chamber resolutions and multiple scattering.  The event
reconstruction checked all possible pairings of the photons to
determine the best $\pi^0$ combinations.

In addition, a large $K_L\rightarrow\pi^0\pi^0\pi^0_{D}$ sample (where
$\pi^0_D$ indicates a Dalitz decay, $\pi^0\rightarrow e^+e^-\gamma$)
was selected from events which had an $e^+e^-$ pair and five CsI
clusters unassociated with tracks.  The
$K_L\rightarrow\pi^0\pi^0\pi^0_{D}$ signal was compared to a Monte
Carlo of this mode incorporating detector acceptances, efficiencies
and resolutions.  Good agreement was observed in numerous kinematic
distributions.  Using the detector simulation of the
$K_L\rightarrow\pi^0\pi^0e^+e^-$ decays, a region of 0.493
GeV/c$^2\leq M_{\pi^0\pi^0e^+e^-}\leq$ 0.501 GeV/c$^2$ and
$p_t^2(e^+e^-\gamma\gamma\gamma\gamma)\leq 1.5\times 10^{-4}$
GeV$^2$/c$^2$ was determined for the $K_L\rightarrow\pi^0\pi^0e^+e^-$
search.  The signal region contained 86\% of the
$K_L\rightarrow\pi^0\pi^0e^+e^-$ decays (using the Sehgal model for
our generated signal events).  Here,
$p_t^2(e^+e^-\gamma\gamma\gamma\gamma)$ was calculated relative to the
vector from the center of the target to the $K_L$ decay vertex.

The most serious background to the $K_L\rightarrow\pi^0\pi^0 e^+e^-$
mode resulted from $K_L\rightarrow\pi^0\pi^0\pi^0$ decays in several
ways: a) one photon from a $\pi^0\rightarrow\gamma\gamma$ decay
converted internally or externally into an $e^+e^-$ pair while one of
the other photons in the decay fails to produce a detectable energy
deposit in the CsI or missed the CsI and the remaining four photons
formed a good $\pi^0\pi^0$ pair; b) two of the six final state photons
converted internally or externally into an $e^+e^-$ pair and an $e^+$
and $e^-$ were lost; c) one of the $\pi^0$ decayed into an $e^+e^-$.
In addition to the backgrounds from $K_L\rightarrow\pi^0\pi^0\pi^0$
decays, $K_L\rightarrow\pi^0\pi^0_D$ decays with an extra accidental
CsI energy cluster contributed a background.  Finally, other
backgrounds due to $K_L\rightarrow\pi^0 e^+e^-\gamma$,
$K_L\rightarrow\pi^0\pi^0\gamma$, and
$K_L\rightarrow\pi^0\pi^0\gamma\gamma$ decays were found to be
negligible due to physics cuts discussed below and to small branching
ratios.

Special cuts were required to eliminate each type of background.  The
background due to $K_L\rightarrow\pi^0\pi^0\pi^0$ decays in which two
photons converted into $e^+e^-$ pairs or one of the $\pi^0$'s decayed
into two $e^+e^-$ pairs was suppressed by rejecting events in which
there were extra track segments upstream or downstream of the magnet.
If a photon converted externally in the upstream spectrometer
material, the resulting $e^+e^-$ did not typically produce two
reconstructed upstream track segments due to the small opening angle.
If one member of the $e^+e^-$ pair had low momentum and was swept out
of spectrometer acceptance by the analysis magnet, no extra track
segments were produced downstream of the magnet either. In these
cases, the invariant mass of the observed charged track and the other
photon from the $\pi^0$ decay was close to the mass of the parent
$\pi^0$ and the event was rejected by cutting on the $\gamma e^{\pm}$
mass.  Specifically, the background from
$K_L\rightarrow\pi^0\pi^0\pi^0$ decays with two photons from different
$\pi^0$ converting to $e^+e^-$ pairs and no extra track segments in
the spectrometer was suppressed by rejecting events with two $e\gamma$
mass combinations with 0.115 GeV/c$^2\leq M_{e\gamma}\leq 0.145$
GeV/c$^2$ and the two remaining photons' mass within 3 MeV/c$^2$ of
the $\pi^0$ mass.

The background due to $K_L\rightarrow\pi^0\pi^0\pi^0$ decays where one
photon converted into an $e^+e^-$ pair and the other photon missed the
calorimeter were rejected by calculating the missing photon energy and
momentum assuming either 1) the missing photon went into the beam hole
and had, in combination with the $e^+e^-$, the $\pi^0$ mass or 2) the
missing photon came from a $K_L$ which had zero $p^2_t$.  If either,
in the first case, $M_{e^+e^-\gamma\gamma\gamma\gamma\gamma}\leq$ 0.51
GeV/c$^2$ and $p^2_t(e^+e^-\gamma\gamma\gamma\gamma\gamma)\leq$ 0.0005
GeV$^2$/c$^2$ or, in the second case, the $e^+e^-$ plus missing gamma
mass was greater than 0.065 GeV/c$^2$, the event was rejected.

To eliminate $K_L\rightarrow\pi^0\pi^0\pi^0$ decays where one photon
converted into an $e^+e^-$ and the other photon was lost due to
coalescing with another photon in the CsI, all observed clusters were
tested to see if any cluster could be due to an overlap of two
photons.  The energy of each observed cluster in turn was divided in
various proportions and the two resulting clusters were used in a fit
of the event to the $K_L\rightarrow\pi^0\pi^0\pi^0_D$ hypothesis.  All
possible combinations of all different clusters with different split
fractions were combined with the other photons and the $e^+e^-$ pair
were considered to see if any combination was consistent with a
$K_L\rightarrow\pi^0\pi^0\pi^0_D$ event.  If the $\pi^0\pi^0\pi^0_{D}$
invariant mass or the vertex $\chi^2$ of a combination was consistent
with a $K_L\rightarrow\pi^0\pi^0\pi^0_{D}$ decay, the event was
rejected.

The background due to $K_L\rightarrow\pi^0\pi^0\pi^0$ decays in which
one $\pi^0$ decayed into an $e^+e^-$ pair (BR($\pi^0\rightarrow
e^+e^-$)= $6.09\times 10^{-8}$~\cite{CP7}) was eliminated by requiring
that $M_{e^+e^-}\leq 0.10$ GeV/c$^2$.  This cut also eliminated the
background from a $K_L\rightarrow\pi^0\pi^0\pi^0$ decay in which one
of the $\pi^0$'s decayed into $e^+e^-e^+e^-$ and one of the $e^+e^-$
pairs had low energy and went undetected.  The background due to
$K_L\rightarrow\pi^0\pi^0_D$ decays with an accidental cluster in the
calorimeter was eliminated by requiring that
$M_{e^+e^-\gamma\gamma\gamma}\leq$ 0.45 GeV/c$^2$.  All
$e^+e^-\gamma\gamma\gamma$ combinations were checked and the mass cut
was applied to the one with the smallest $\chi^2$ for a
$K_L\rightarrow\pi^0\pi^0_D$ decay.

\begin{figure} [h!]
  \epsfxsize=7.0cm \epsfbox{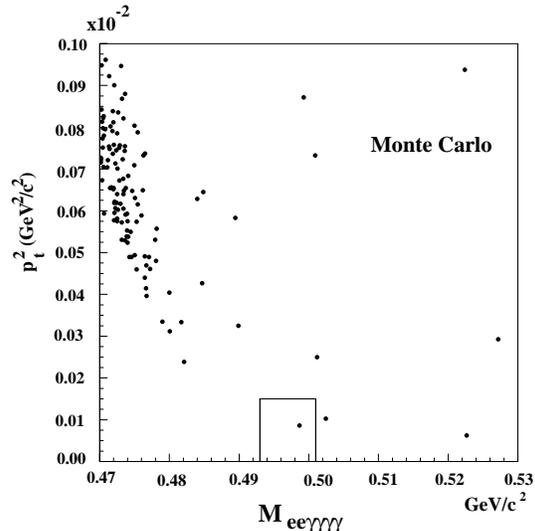}
  \caption{Monte Carlo simulation of backgrounds to
  $K_L\rightarrow\pi^0\pi^0e^+e^-$ $p_t^2$
  vs. $M_{e^+e^-\gamma\gamma\gamma\gamma}$. The total number of
  background events in the plot is equivalent to three times the
  actual integrated flux of the KTeV data.}  \label{fig:backgrounds}
\end{figure}

After these cuts, the major remaining background, as determined from
Monte Carlo simulations, was due to
$K_L\rightarrow\pi^0\pi^0\pi^0_{D}$ decays. All other backgrounds were
negligible. The contributions from all types of
$K_L\rightarrow\pi^0\pi^0\pi^0_{D}$ backgrounds are shown in
Fig.~\ref{fig:backgrounds} as a function of $p^2_t$ and
$M_{e^+e^-\gamma\gamma\gamma\gamma}$ for the equivalent of three E799
data sets after all cuts.  One event is observed in the signal region
and the backgrounds outside the signal region are similar in
distribution and magnitude to the background observed in data when
normalized to the E799 integrated flux. One event from the background
Monte Carlo of three E799 data sets leads to an estimated background
to the $K_L\rightarrow\pi^0\pi^0\pi^0_{D}$ signal mode of
$0.33^{+0.56}_{-0.21}$ events.

\begin{figure} [h!]
  \epsfxsize=7.0cm \epsfbox{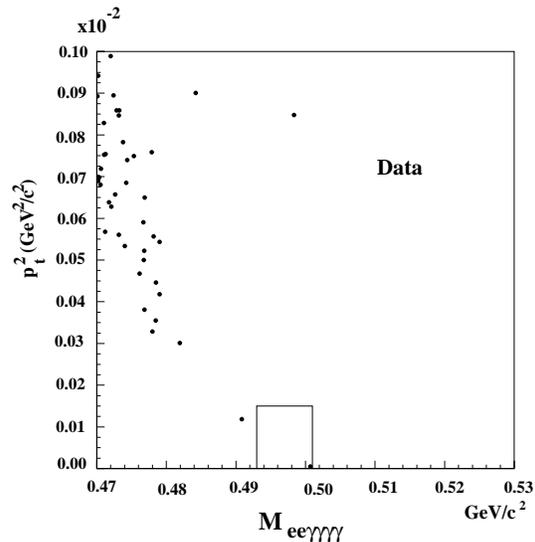}
  \caption{Events from the experimental data surviving all criteria
  for $K_L\rightarrow\pi^0\pi^0e^+e^-$ decays vs. $p_t^2$ and
  $M_{e^+e^-\gamma\gamma\gamma\gamma}$.}  \label{fig:data}
\end{figure}

\begin{figure} [h!]
  \epsfxsize=8.5cm \epsfbox{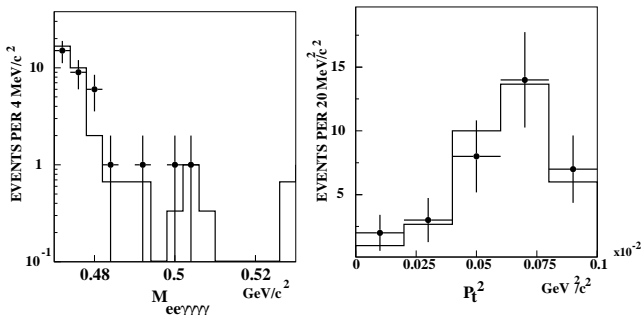}
  \caption{Comparison of $K_L\rightarrow\pi^0\pi^0e^+e^-$ data (dots)
  and Monte Carlo (histogram) background distributions for
  a)$M_{e^+e^-\gamma\gamma\gamma\gamma}$ and b) $p_t^2$ distributions
  of events surviving cuts.}  \label{fig:proj}
\end{figure}

Using the vertex fit, the mass of M$_{ee\gamma\gamma\gamma\gamma}$ and
$p_t^2$ of the $e^+e^-\gamma\gamma\gamma\gamma$ events from the
triggers requiring two charged track and six cluster were calculated
and the cuts mentioned above determined from Monte Carlo simulations
of the signal and various backgrounds or from the
$K_L\rightarrow\pi^0\pi^0\pi^0_{D}$ data events were applied.
Figure~\ref{fig:data} shows the data remaining after applying the
cuts.  One event was observed in the signal box. Shown in
Fig.~\ref{fig:proj} are the comparisons of the data and Monte Carlo
simulations of the events vs. $p_t^2$ and
$M_{e^+e^-\gamma\gamma\gamma\gamma}$.

An overall product of the acceptance and efficiency of 0.25\% for the
signal mode was obtained from a Monte Carlo simulation of
$K_L\rightarrow\pi^0\pi^0e^+e^-$ mode using the model of
Ref.~\cite{CP3} with a chiral perturbation parameter $w_s$=
0.46~\cite{CP5}. Since estimations of $w_s$ from exisiting
experimental data are indirect and model dependent, we have studied
the variation of our acceptance as a function of $w_s$. We found that
a simulation using a $w_s$= 0.9 gives a minimum acceptance of 0.205\%
in the model of Ref.~\cite{CP3} (and a maximum upper limit for
$K_L\rightarrow\pi^0\pi^0e^+e^-$ branching ratio).  By comparison, a
simulation using phase space for the decay gave an acceptance of
0.29\%. We have used the lower acceptance 0.205\% for calculating the
single event sensitivity using in determining a
$K_L\rightarrow\pi^0\pi^0e^+e^-$ branching ratio upper limit.

\begin{figure} [h!]
  \epsfxsize=8.0cm \epsfbox{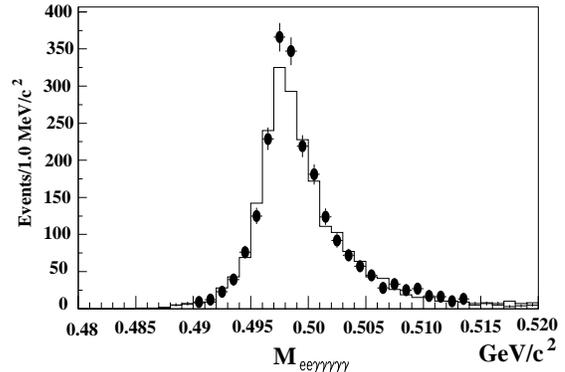}
  \caption{The mass distribution of
  $K_L\rightarrow\pi^0\pi^0\pi^0_{D}$ in which the photon from the
  $\pi^0\rightarrow e^+e^-\gamma$ decay passes into one of beam holes
  in the CsI calorimeter and is reconstructed assuming that its
  trajectory passes through the center of the given beam hole.  The
  solid dots are the data and the histogram is the Monte Carlo
  simulation of these events.}  \label{fig:3pi0D}
\end{figure}

$K_L\rightarrow\pi^0\pi^0\pi^0_{D}$ events with a missing $\gamma$
down the beam hole (which could be reconstructed as
$K_L\rightarrow\pi^0\pi^0\pi^0_{D}$ using the technique for treating
beam hole photons discussed above) were used as a normalization mode
since they had the same trigger and analysis cuts as the signal mode.
Figure~\ref{fig:3pi0D} shows the mass distribution of these events
compared to the Monte Carlo simulation.  A distinct $K_L$ peak is
reconstructed for the data with a resolution matching the simulation
of this type of events. The disagreement between the level of data and
Monte Carlo of the normalization events has been folded into the
systematic error in the single event sensitivity.

A single event sensitivity of $1.71\pm0.06(stat)\pm0.19(syst)\times
10^{-9}$ was calculated using the ratio of the
$K_L\rightarrow\pi^0\pi^0e^+e^-$ signal efficiency to the
$K_L\rightarrow\pi^0\pi^0\pi^0_{D}$ normalization sample efficiency,
the $K_L\rightarrow\pi^0\pi^0\pi^0_{D}$ branching ratio of $7.59\times
10^{-3}$~\cite{pdg} and the observed 2234
$K_L\rightarrow\pi^0\pi^0\pi^0_{D}$ normalization events of the type
described in the previous paragraph.  The observation of one event in
the signal box resulted in an upper limit for
BR($K_L\rightarrow\pi^0\pi^0e^+e^-$) of $6.6\times 10^{-9}$ at the
90\% confidence level.  Using the ratio of the
BR$(K_L\rightarrow\pi^0\pi^0\gamma)$ to
BR$(K_L\rightarrow\pi^0\pi^0e^+e^-)$ of $\approx$50~\cite{CP3}, our
result is a factor of 10 more sensitive than the present upper limit
of $5.6\times 10^{-6}$ from a direct search~\cite{Barr} for
$K_L\rightarrow\pi^0\pi^0\gamma$.

In conclusion, we have made the first attempt to detect the
$K_L\rightarrow\pi^0\pi^0e^+e^-$ decay mode.  We have established an
upper limit of $6.6\times 10^{-9}$ at the 90\% confidence level (using
a $w_s$= 0.9 in the model of Ref.~\cite{CP3}) which excludes the
vector dominance model of Ref.~\cite{CP4} but does not quite reach the
level of expected branching ratios predicted by the more recent models
of Refs.~\cite{CP3,CP5}.

We gratefully acknowledge the support and effort of the Fermilab staff
and the technical staffs of the participating institutions for their
vital contributions.  This work was supported in part by the U.S.
Department of Energy, The National Science Foundation and The Ministry
of Education and Science of Japan.

\vspace{.1in}

$^{\dagger}$ To whom correspondence should be addressed. \\
$^{*}$ Permanent address C.P.P. Marseille/C.N.R.S., France \\
$^{**}$Permanent address Univ. of S\~{a}o Paulo, S\~{a}o Paulo, Brazil \\

\end{document}